\documentstyle{mn}\twocolumn

\newif\ifAMStwofonts

\def\mincir{\raise -2.truept\hbox{\rlap{\hbox{$\sim$}}\raise5.truept \hbox{$<$}\ }}
\def\mincireq{\hbox{\raise0.5ex\hbox{$<\lower1.06ex\hbox{$\kern-1.07em{\sim}$}$}}}
\def\magcir{\raise-2.truept\hbox{\rlap{\hbox{$\sim$}}\raise5.truept \hbox{$>$}\ }}

\title{Non-thermal emission in lobes of radio galaxies:\,\,  \\ 
III. 3C\,98, Pictor\,A, DA\,240, Cygnus\,A, 3C\,326, and 3C\,236}

\author[Persic \& Rephaeli]
       {Massimo Persic$^{1,2,3}$, 
        Yoel Rephaeli$^{4,5}$\\
        $^1$INAF-Trieste Astronomical Observatory, via G.B.\,Tiepolo 11, I-34100 Trieste, Italy \\
        $^2$INFN-Trieste, via A.\,Valerio 2, I-34127 Trieste, Italy \\
        $^3$Physics \& Astronomy Dept., Bologna University, via P.\,Gobetti 93/2, I-40129 Bologna, Italy \\
        $^4$School of Physics \& Astronomy, Tel Aviv University, Tel Aviv 69978, Israel \\
        $^5$Center for Astrophysics and Space Sciences, University of California at 
		San Diego, La Jolla, CA 92093, USA} 
\date{Accepted ... ... ... ... ;
      Received ... ... ... ... ;
      in original form ... ... ... ...}

\pubyear{2019}

\begin{document}

\maketitle

\label{firstpage}

\begin{abstract}
Recent analyses of the broad spectral energy distributions (SED) of extensive lobes of local radio-galaxies have 
confirmed the leptonic origin of their {\it Fermi}/LAT $\gamma$-ray emission, significantly constraining the level 
of hadronic contribution. SED of distant ($D_L>125$ Mpc) radio-galaxy lobes are currently limited to the radio and 
X-ray bands, hence give no information on the presence of non-thermal (NT) protons but are adequate to describe the 
properties of NT electrons. 
Modeling lobe radio and X-ray emission in 3C\,98, Pictor\,A, DA\,240, Cygnus\,A, 3C\,326, and 3C\,236, we fully 
determine the properties of intra-lobe NT electrons and estimate the level of the related $\gamma$-ray emission 
from Compton scattering of the electrons off the superposed Cosmic Microwave Background, Extragalactic Background 
Light, and source-specific radiation fields. 
\end{abstract}


\maketitle\markboth{Persic \& Rephaeli: NT emission in giant radio-galaxy lobes}{}

\section{Introduction}

Measurements of non-thermal (NT) emission from the extended lobes of radio galaxies provide a 
tangible basis for 
modeling of the spectral distributions of energetic particles in these environments. Sampling the spectral energy 
distributions (SED), even with only limited spatial information, yields valuable insight on the emitting electrons 
and possibly also on energetic protons whose {\it p--p} interactions in the ambient lobe plasma and ensuing 
$\pi^0$-decay could yield detectable $\magcir 0.1$ GeV emission. In addition to the intrinsic interest in physical 
conditions in radio lobes, modeling energetic particles and their emission processes in these environments can 
yield clues to the origin of NT particles in galaxy clusters.

Currently available spectral radio, X-ray, and $\gamma$-ray measurements of the lobes of four radio galaxies have 
provided an adequate basis for determining the emission processes, the SED of the NT emitting particles, and the 
mean value of the magnetic field when the superposed photon fields in the lobe region are properly accounted for 
(Fornax\,A: Persic \& Rephaeli 2019a, hereafter Paper I; Centaurus\,A, Centaurus\,B, and NGC\,6251: Persic \& 
Rephaeli 2019b, hereafter Paper II). These SED analyses have confirmed the leptonic origin of the measured lobe 
{\it Fermi}/LAT $\gamma$-ray emission, constraining the level of hadronic contribution to within several percent. 

In the present paper we extend our SED analysis of radio-galaxy lobes to six relatively distant ($D_L > 125$ Mpc) 
sources with no available $>$0.1 GeV {\it Fermi}/LAT fluxes: 3C\,98, Pictor\,A, DA\,240, Cygnus\,A, 3C\,326, and 
3C\,236. Lacking $\gamma$-ray data, the available spectral measurements enable determination of the electron 
spectral distribution, but can not directly constrain NT proton yields. Since the results presented here are based 
essentially on an identical treatment to that in Papers\,I and II, our discussion will be limited only to the most 
relevant observational data and to the results of our spectral modeling. In Section 2 we summarize the observational 
data and estimates of the radiation field densities in the lobes of the six galaxies. Results of the fitted SED models 
are detailed in Section 3 and discussed in Section 4. Our main conclusions are summarized in Section 5.

\section{Emission and radiation fields in the lobes}

The six radio galaxies included in this analysis have elliptical hosts and a double-lobe morphology, with (usually) 
a flat radio lobe surface brightness. In this section we briefly discuss the sources, observations of NT emission, 
the lobe superposed radiation fields.

\subsection{Observations of NT emission}

3C\,98 ($z=0.0306$; luminosity distance $D_L = 126$\,Mpc) shows a double-lobe morphology with a flat surface brightness 
distribution and a total angular extent of $5^\prime \times 2^\prime$ (Leahy et al. 1997). Broad-band total flux 
densities from a variety of radio telescopes, compiled in the NASA Extragalactic Database (NED), are listed in Table 1. 
NT X-ray emission, detected from each lobe, is interpreted as arising from Compton scattering of Cosmic Microwave Background 
(CMB) photons off radio emitting electrons (Isobe et al. 2005: {\it XMM-Newton} data). We assume a lobe radius $r=30$ kpc 
and a galaxy to nearest lobe boundary distance $d=26.5$ (see Isobe et al. 2005). 

Pictor\,A ($z=0.0342$; $D_L=149$ Mpc) shows two symmetrical dim radio lobes with a bright compact hot spot on the far edge of 
the W lobe (connected with the radio nucleus by a faint jet) and two fainter, less compact spots on the E lobe (Perley et al. 
1997). NT X-ray emission from the lobes, detected by Grandi et al. (2003, {\it XMM-Newton} data) and confirmed in its spatial 
extension by Hardcastle \& Croston (2005, {\it Chandra} data; see also Hardcastle et al. 2016) and Migliori et al. (2007, {\it 
XMM-Newton} data), is interpreted as Compton/CMB radiation. The detected 0.2-200 GeV $\gamma$-ray emission is attributed to 
the jet (Brown \& Adams 2012; {\it Fermi}/LAT data), based on its variability timescale ($\mincir$1\,yr) and its incompatibility 
with the (well resolved) radio and X-ray 
emission 
within a synchrotron self-Compton SED model of a prominent compact hotspot 
in the Western lobe. Lobe NT SED data are reported in Table 2. We assume $r=65$ kpc and $d=15$ kpc (Perley et al. 1997). 

DA\,240 ($z=0.03566$; $D_L = 152$\,Mpc; $\theta \sim 35^\prime$), 
3C\,326 ($z=0.0895$; $D_L = 395$\,Mpc; $\theta \sim 20^\prime$), and 
3C\,236 ($z=0.1005$; $D_L = 449$\,Mpc; angular extent $\theta \sim 40^\prime$)
have 0.326-10.6 GHz data from the entire sources and from their individual components (Mack et al. 1997). 
NT X-ray emission, 
detected from the E lobe of DA\,240 and the W lobes of 3C\,326 and 3C\,236, is interpreted as Compton/CMB radiation 
(3C\,236: Isobe \& Koyama 2015; 3C\,326: Isobe et al. 2009; DA\,240: Isobe et al. 2011 -- {\it Suzaku} data). Lobe SED data 
for the three sources are reported in Tables 3-5. We assume $r=400$ kpc and $d=850$ kpc for 3C\,236 (Isobe \& Koyama 2014), 
$r=225$ kpc and $d=515$ kpc for 3C\,326 (see Ogle et al. 2007 and Isobe et al. 2009), and $r=268.5$ kpc and $d=190$ kpc for 
DA\,240 (Isobe et al. 2011). 

Cygnus\,A ($z=0.0561$; $D_L = 237$\,Mpc), with multi-frequency radio (VLA) and X-ray ({\it Chandra}) emissions available 
for both lobes separately (Yaji et al. 2010; de Vries et al. 2018), is the brightest radio galaxy in the sky (Birzan et 
al. 2004). NT X-ray emission from both lobes is interpreted as Compton scattering off the CMB and from synchrotron radiation 
from the lobes (Yaji et al. 2010; de Vries et al. 2018). Within the relative paucity of  data, the lobe SEDs appear to be 
different, with that of the E-lobe steeper/flatter than that of the W-lobe at radio/X-ray frequencies (Yaji et al. 2010; 
de Vries et al. 2018). Lobe NT SED data are reported in Table\,6. We assume $r=16.3$ kpc and $d=61.3$ kpc (Yaji et al. 
2010).  

\begin{table*}

\caption[] {Emission from the lobes of 3C\,98.}

\begin{tabular}{ c  c  c  c  c  c  c  c  c  c  c }
\hline
\hline

Frequency&   Flux Density  & &  Frequency   & Flux Density   & &Frequency     &   Flux Density  & & Frequency     &Flux Density \\
Log($\nu$/Hz) &  [Jy]      & &Log($\nu$/Hz) &  [Jy]          & &Log($\nu$/Hz) &  [Jy]           & & Log($\nu$/Hz) &  [Jy]       \\
\hline

~\,7.167 & $480 \pm 110$   & &  ~\,7.934  &$ 94.3 \pm 2.2$   & &  ~\,8.875  & $16.7 \pm 0.3$    & &  ~\,9.699  & $ 3.29         $  \\
~\,7.223 & $420 \pm  80$   & &  ~\,8.204  &$ 49.1 \pm 6.4$   & &  ~\,8.875  & $16.0 \pm 0.8$    & &  ~\,9.699  & $ 4.97 \pm 0.25$  \\
~\,7.301 & $390 \pm  66$   & &  ~\,8.250  &$ 44.0 \pm -  $   & &  ~\,9.146  & $10.2 \pm 0.5$    & &  ~\,9.699  & $ 4.94 \pm 0.25$  \\
~\,7.301 & $289 \pm  29$   & &  ~\,8.250  &$ 48.8 \pm 3.9$   & &  ~\,9.146  & $10.3 \pm 0.3$    & &  ~\,9.700  & $ 4.73 \pm 0.31$  \\
~\,7.347 & $312 \pm 43.3$  & &  ~\,8.250  &$ 51.4 \pm 2.6$   & &  ~\,9.146  & $ 9.75 \pm 0.2$   & &  ~\,9.700  & $ 3.39 \pm 0.12$  \\
~\,7.348 & $312 \pm  30$   & &  ~\,8.250  &$ 35.5 \pm 2.8$   & &  ~\,9.146  & $ 9.9 \pm 0.5$    & &  ~\,9.903  & $ 3.08 \pm 0.07$  \\
~\,7.398 & $260 \pm  52$   & &  ~\,8.250  &$ 50.6 \pm 2.5$   & &  ~\,9.146  & $11.1        $    & &    10.029  & $ 2.82 \pm 0.12$  \\
~\,7.398 & $285 \pm  34$   & &  ~\,8.250  &$ 47.2 \pm 4.7$   & &  ~\,9.149  & $12.0 \pm 0.3$    & &    10.029  & $ 3.01 \pm 0.13$  \\
~\,7.420 & $215 \pm  18$   & &  ~\,8.502  &$ 29.7 \pm 1.1$   & &  ~\,9.423  & $ 7.30 \pm 0.20$  & &    10.362  & $ 1.35 \pm 0.20$  \\
~\,7.420 & $218 \pm  17$   & &  ~\,8.611  &$ 26.9        $   & &  ~\,9.431  & $ 7.01 \pm 0.35$  & &    10.362  & $ 1.20 \pm 0.12$  \\
~\,7.580 & $147 \pm  22$   & &  ~\,8.611  &$ 21.3 \pm 0.9$   & &  ~\,9.431  & $ 7.09 \pm 0.10$  & &    10.519  & $ 1.10 \pm 0.10$  \\
~\,7.580 & $160 \pm 7.2$   & &  ~\,8.611  &$ 24.2 \pm 4.6$   & &  ~\,9.431  & $ 7.17 \pm 0.35$  & &    10.613  & $ 0.7  \pm 0.2 $  \\
~\,7.580 & $173 \pm 8.7$   & &  ~\,8.611  &$ 27.5        $   & &  ~\,9.431  & $ 6.10        $   & &    10.795  & $ 0.3  \pm 0.3 $  \\
~\,7.778 & $132 \pm  30$   & &  ~\,8.670  &$ 26.7 \pm 1.5$   & &  ~\,9.679  & $ 2.70        $   & &    10.973  & $ 0.1  \pm 0.5 $  \\
~\,7.869 & $ 98 \pm 1.1$   & &  ~\,8.803  &$ 20.9 \pm 0.4$   & &  ~\,9.686  & $ 3.13 \pm 0.43$  & &    17.383  & $(15.6 \pm 2.5)$ E$-$9 \\
~\,7.903 & $ 90 \pm 13 $   & &  ~\,8.875  &$ 16.7 \pm 0.8$   & &  ~\,9.686  & $ 5.05 \pm 0.76$  & &            &                   \\
~\,7.903 & $ 83 \pm -  $   & &  ~\,8.875  &$ 17.7 \pm 0.3$   & &  ~\,9.699  & $ 4.93 \pm 0.12$  & &            &                   \\ 

\hline
\end{tabular}

\begin{flushleft}
\noindent
Data: NED (radio), Isobe et al. (2005; X-rays). Unspecified flux errors are assumed to be at the 10\% level.
\end{flushleft}

\end{table*}

\begin{table}
\caption[] {Emission from the lobes of Pictor\,A.}

\begin{tabular}{ c  c  c  c  c  c  c  c  c  c  c  }
\hline
\hline

 & Frequency     &  &  &    East Lobe         &  &  &     West Lobe       &  \\
 & Log($\nu$/Hz) &  &  &      [Jy]            &  &  &       [Jy]          &  \\
\hline

 &  ~\,7.869     &  &  &   $237$              &  &  &    $213$            &  \\
 &  ~\,8.515     &  &  &  $86.3$              &  &  &   $94.3$            &  \\
 &  ~\,9.166     &  &  &  $27.4$              &  &  &   $32.9$            &  \\
 &  ~\,9.699     &  &  &  $8.77$              &  &  &   $12.0$            &  \\
 &    17.383     &  &  &   $(55 \pm 2)$ E$-$9 &  &  &  $(55 \pm 2)$ E$-$9 &  \\
\hline
\end{tabular}

\begin{flushleft}
\noindent
Data: Perley et al. (1997; radio), Hardcastle \& Croston (2005; X-rays -- see also Hardcastle et al. 2016). 
Errors on radio fluxes are assumed to be at the 10\% level.
\end{flushleft}

\end{table}

\begin{table}
\caption[] {Emission from the W lobe of DA\,240.}

\begin{tabular}{ c  c  c  c  c  }
\hline
\hline

Frequency     &   Flux Density  & & Frequency     &Flux Density \\
Log($\nu$/Hz) &  [mJy]           & & Log($\nu$/Hz) &  [mJy]       \\
\hline

 ~\,8.513  & $10299.0 \pm 120.4$   & &  ~\,9.677  & $ 1186.6 \pm 23.4   $  \\
 ~\,8.785  & $\,5688.7 \pm 78.1$   & &    10.023  & $\,749.3 \pm 27.8   $  \\
 ~\,9.431  & $\,1809.6 \pm 31.2$   & &    17.383  & $(51.5 \pm 7.0)$ E$-$6  \\

\hline
\end{tabular}

\begin{flushleft}
\noindent
Data: Mack et al. (1997; radio), Isobe et al. (2011; X-rays). 
\end{flushleft}

\end{table}

\begin{table}
\caption[] {Emission from the lobes of Cygnus\,A.}

\begin{tabular}{ c  c  c  c  c  c  c  c  c  }
\hline
\hline

 & Frequency     &  &  &     East Lobe         &  &  &     West Lobe        &  \\
 & Log($\nu$/Hz) &  &  &       [Jy]            &  &  &       [Jy]           &  \\
\hline

 &  ~\,9.129     &  &  &  $594 \pm 36$         &  &  &  $429 \pm 29$        &  \\
 &  ~\,9.231     &  &  &  $463 \pm 29$         &  &  &  $357 \pm 24$        &  \\
 &  ~\,9.656     &  &  &  $129 \pm 10$         &  &  &  $122 \pm 9$         &  \\
 &  ~\,9.699     &  &  &  $115 \pm 9$          &  &  &  $108 \pm 9$         &  \\
 &    17.383     &  &  &   $(71 \pm 10)$ E$-$9 &  &  &  $(50 \pm 13)$ E$-$9 &  \\
\hline
\end{tabular}

\begin{flushleft}
\noindent
Data: Yaji et al. (2010; radio), de Vries et al. (2018; X-rays). 
\end{flushleft}

\end{table}

\begin{table}
\caption[] {Emission from the W lobe of 3C\,326.}

\begin{tabular}{ c  c  c  c  c  }
\hline
\hline

Frequency     &   Flux Density  & & Frequency     &Flux Density \\
Log($\nu$/Hz) &  [mJy]           & & Log($\nu$/Hz) &  [mJy]       \\
\hline

 ~\,8.513  & $1534.6 \pm 22.9$  & &  ~\,9.677  & $ 232.1 \pm 2.4   $  \\
 ~\,8.785  & $982.5 \pm 18.8$   & &    10.023  & $ 114.1 \pm 5.6   $  \\
 ~\,9.431  & $933.6 \pm 43.7$   & &    17.383  & $(19.4 \pm 4.4)$ E$-$6  \\

\hline
\end{tabular}

\begin{flushleft}
\noindent
Data: Mack et al. (1977; radio), Isobe et al. (2009; X-rays). 
\end{flushleft}

\end{table}

\begin{table}
\caption[] {Emission from the W lobe of 3C\,236.}

\begin{tabular}{ c  c  c  c  c  }
\hline
\hline

Frequency     &   Flux Density  & & Frequency     &Flux Density \\
Log($\nu$/Hz) &  [mJy]           & & Log($\nu$/Hz) &  [mJy]       \\
\hline

 ~\,8.513  & $588.1 \pm 22.3$    & &  ~\,9.677  & $ 36.4 \pm 14.4   $  \\
 ~\,8.785  & $348.6 \pm 18.3$    & &    10.023  & $ 40.3 \pm 16.3   $  \\
 ~\,9.431  & $153.0 \pm 18.3$    & &    17.383  & $(12.3 \pm 2.8)$ E$-$6  \\

\hline
\end{tabular}

\begin{flushleft}
\noindent
Data: Mack et al. (1997; radio), Isobe \& Koyama (2014; X-rays). 
\end{flushleft}

\end{table}

\subsection{Radiation fields}

A precise determination of the ambient photon fields in the lobes is needed to predict X/$\gamma$-ray emission 
from Compton scattering of radio-emitting electrons. Radiation fields in the lobes include cosmic and local 
components.

Cosmic radiation fields include the CMB and the Extragalactic Background Light (EBL). The CMB is a pure Planckian with 
$T_{\rm CMB} = 2.725$\,K and energy density $u_{\rm CMB} = 0.25\, (1+z)^{4}$\,eV\,cm$^{-3}$ (e.g., Dermer \& 
Menon 2009). The EBL, originating from direct and dust-reprocessed starlight integrated over the star formation 
history of the universe (e.g., Franceschini \& Rodighiero 2017; Acciari et al. 2019), can be represented as a 
suitable combination of diluted Planckians (see Paper II). 

Local radiation fields (Galaxy Foreground Light, GFL) arise from the elliptical host galaxies, whose SEDs usually 
show two thermal humps, IR and optical. 
\smallskip

\noindent
{\it (i)} 3C\,98's host galaxy has a bolometric IR luminosity $L_{\rm IR} < 4.1 \cdot 10^{43}$ erg s$^{-1}$, as 
implied by {\it IRAS} flux densities at 12, 25, 60 and 100$\mu$m (Golombek et al. 1988)
\footnote{
The total IR ($8-1000\,m$m) flux is $f_{\rm IR}=1.8 \cdot 10^{-11} (13.48\,f_{12} + 5.16\,f_{25} + 2.58\,f_{60} + 
f_{100})$\, erg\,cm$^{-2}$s$^{-1}$ (Sanders \& Mirabel 1996), where $f_{12}$, $f_{25}$, $f_{60}$, $f_{100}$ are the 
{\it IRAS} flux densities at the indicated wavelengths (in $\mu$n).}
. The optical bolometric luminosity $L_{\rm opt} \sim 2.3 \cdot 10^{44}$\,erg\,s$^{-1}$ is derived from $V = 14.12$ mag 
(Smith \& Heckman 1989a) applying the bolometric correction ($BC_{\rm V} = -0.85$, Buzzoni et al. 2006), so that $m_{\rm 
opt}^{\rm bol} = V + BC_V$ (see Paper I). 
\smallskip

\noindent
{\it (ii)} Pictor\,A's host has $L_{\rm IR} \sim 1.3 \cdot 10^{44}$ erg s$^{-1}$, estimated from the $f_{12}$, 
$f_{25}$, $f_{60}$ (Singh et al. 1990) and $f_{100}$ (inferred from $f_{60}$, see Lisenfeld et al. 2007) {\it 
IRAS} flux densities; and $L_{\rm opt} \sim 1.2 \cdot 10^{44}$ erg s$^{-1}$, estimated from $B = 15.95$, (B-V)$ 
= 0.73$ (from NED) applying the bolometric correction. 
\smallskip

\noindent
{\it (iii)} DA\,240's host has $L_{\rm opt} \sim 5 \cdot 10^{45}$ erg s$^{-1}$, estimated from $R = 10.7$ mag 
(Peng et al. 2004) converted to V mag using (B-V)=0.90 
\footnote{ www.aerith.net/astro/color$_{_{-}}$conversion.html }
and applying the bolometric correction. $L_{\rm IR} \sim 6 \cdot 10^{45}$ erg s$^{-1}$ is estimated from $L_B$ 
through Bregman et al.'s (1998) FIR-B relation assuming $L_{\rm IR} \sim 2\, L_{\rm FIR}$ (Persic \& Rephaeli 2007). 
\smallskip

\noindent
{\it (iv)} Cygnus\,A's host has $L_{\rm IR} \sim 1.9 \cdot 10^{45}$ erg s$^{-1}$ (Golombek et al. 1988), and 
$L_{\rm opt} \sim 1.5 \cdot 10^{45}$\,erg\,s$^{-1}$ estimated from $V = 13.46$ (total, extinction corrected; 
Smith \& Heckman 1989b) applying the bolometric correction. The IR emission is (Privon et al. 2012) mostly 
($\sim$90\%) torus-reprocessed radiation by an optically obscured ($A_V > 50$ mag, Imanishi \& Ueno 2000) AGN 
(Antonucci et al. 1994; Carilli et al. 2019) with additional contributions from a dust-enshrouded starburst 
($\magcir$6\%) and a synchrotron jet ($\mincir$4\%).
\smallskip

\noindent
{\it (v)} 3C\,326's host has $L_{\rm IR} \sim 1.3 \cdot 10^{43}$ erg s$^{-1}$ (Ogle et al. 2007), and $L_{\rm opt} 
\sim 1.2 \cdot 10^{44}$ erg s$^{-1}$ derived from $L_B$ which is deduced from $L_{\rm FIR} \sim 1/2\, L_{\rm IR}$ 
(e.g. Persic \& Rephaeli 2007) through a FIR-B relation (Bregman et al. 1998). 
\smallskip

\noindent
{\it (vi)} 3C\,236's central galaxy has $L_{\rm IR} < 4.1 \cdot 10^{43}$ erg s$^{-1}$ (see Golombek et al. 1988), and 
$L_{\rm opt} \sim 6.9 \cdot 10^{44}$ erg s$^{-1}$ derived from $V = 15.72$ mag (Smith \& Heckman 1989b) applying the 
bolometric correction.
\smallskip

These IR and optical parameters allow us to model the GFL; in our calculations we take $T_{\rm gal,\,OPT} 
= 2900$\,K and $T_{\rm gal,\,IR} = 29$\,K (see PR19). The lobe X-ray data are spatial averages, so we 
correspondingly compute volume-averaged Compton/GFL yields, based on the fact that lobe radii and projected 
distances (from central galaxies) are much larger than the corresponding central-galaxy radii
\footnote{
Effective radii estimated from B-luminosities (Romanishin 1986), or isophotal radii (Smith \& Heckman 1989a). 
}
, we treat galaxies as point sources (see Paper II).

\section{Modelling lobe NT emission}

Radio emission in the lobes is by electron synchrotron in a disordered magnetic field whose mean value $B$ 
is taken to be spatially uniform, and X-$\gamma$ emission is by electron Compton scattering off the CMB and 
optical radiation fields. The calculations of the emissivities from these processes are standard (see Paper I). 
Assuming steady state, the electron energy distribution (EED) is assumed to be a time independent, spatially 
isotropic, truncated-PL distribution in the electron Lorentz factor, $N_e(\gamma) = N_{e0}\, \gamma^{-q_e}$ 
in the interval $[\gamma_{min},\, \gamma_{max}$], with a finite $\gamma_{max}$. 

In all cases but Cyg\,A, the photoelectrically absorbed
\footnote{
3C\,98:  $N_H = 1.17 \cdot 10^{21}$ cm$^2$ (Isobe et al. 2005); 
3C\,236: $N_H = 0.93 \cdot 10^{20}$ cm$^2$ (Hardcastle et al. 2016); 
3C\,326: $N_H = 3.84 \cdot 10^{20}$ cm$^2$ (Dickey \& Lockman 1990); 
DA\,240: $N_H = 0.49 \cdot 10^{21}$ cm$^2$ (Isobe et al. 2011); 
Cygnus\,A: $N_H = 0.31 \cdot 10^{22}$ cm$^2$ (De Vries et al. 2018); 
Pictor\,A: $N_H = 0.42 \cdot 10^{21}$ cm$^2$ (Hardcastle \& Croston 2005).
}
1\,keV flux density is used to determine $N_{e0}$ assuming the emission is Compton/CMB. Cyg\,A is a 
notable exception because, due to its much higher magnetic field (than the other sources studied here, 
and in Papers I and II), as revealed by its very high radio to X-ray emission ratio, the synchrotron 
energy density in its lobes ($u_s= 0.55$ eV cm$^{-3}$, in the range $10^{5}$-$10^{11}$ Hz) exceeds the 
CMB energy density, such that synchrotron-self-Compton (SSC) radiation contributes to the 1\,keV flux 
even more than that of Compton/CMB. In calculating the SSC yield (see Eq.\,11 in Paper I) we used an 
analytical expression for the synchrotron photon density field (deduced from a fit to the radio 
synchrotron spectrum
\footnote{
The synchrotron emissivity we used to fit the radio flux spectrum is analytically given by $\eta(\nu) = A_s \, 
(\nu / \nu_\star)^{-\alpha} \, e^{-(\nu / \nu_\star)}$ erg s$^{-1}$ cm$^{-3}$ Hz$^{-1}$, with (for total lobe 
emission) $A_s = 10^{-34.65}$ and $\nu_\star = 7$\,GHz.
}), 
$n_{s}(\epsilon) = N_{s,0}\, (\epsilon/ \epsilon_0)^{-(\alpha+1)}\, e^{-\epsilon/ \epsilon_0}$ cm$^{-3}$ 
erg$^{-1}$, with $N_{s,0} = (r_s/c)\,A_s$ and $\epsilon_0$ the photon energy corresponding to 7\,GHz 
(combined lobes), 6\,GHz (E lobe), and 9\,GHz (W lobe) and $\alpha = (q_e - 1)/2 = 0.75$. Given $n_{s}
(\epsilon)$ and $n_{\rm CMB}(\epsilon)$, we determined $N_{e0}$ for Cyg\,A by fitting the combination of 
predicted SSC and Compton/CMB emissions to the measured 1\,keV flux. 

Matching the synchrotron prediction to the radio spectrum yields $q_e$ (from its featureless PL portion) 
and $\gamma_{\rm max}$ from the (hint of) spectral turnover at high energies. The minimum electron energy, 
$\gamma_{\rm min}$, can be estimated directly from the 1\,keV Compton/CMB data: $\gamma_{\rm min} = 100$, 
corresponds to the transition between Coulomb and synchrotron-Compton losses (Paper II). With the electron 
spectrum fully specified, normalization of the predicted synchrotron spectral flux to the radio measurements 
yields $B$. The model SED and data are shown in Fig.\,1. 

\begin{figure*}
\vspace{12.5cm}
\includegraphics{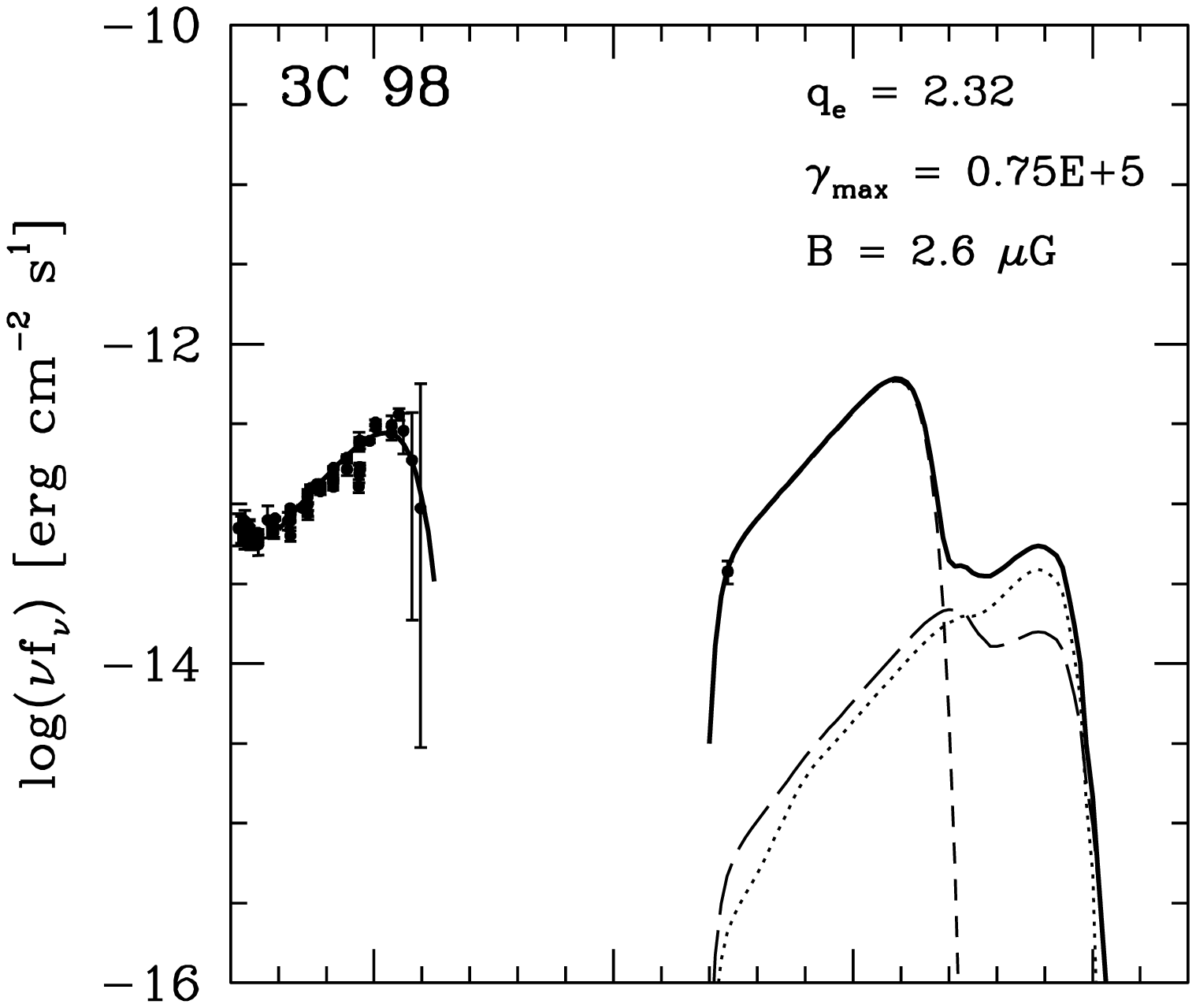}
\includegraphics{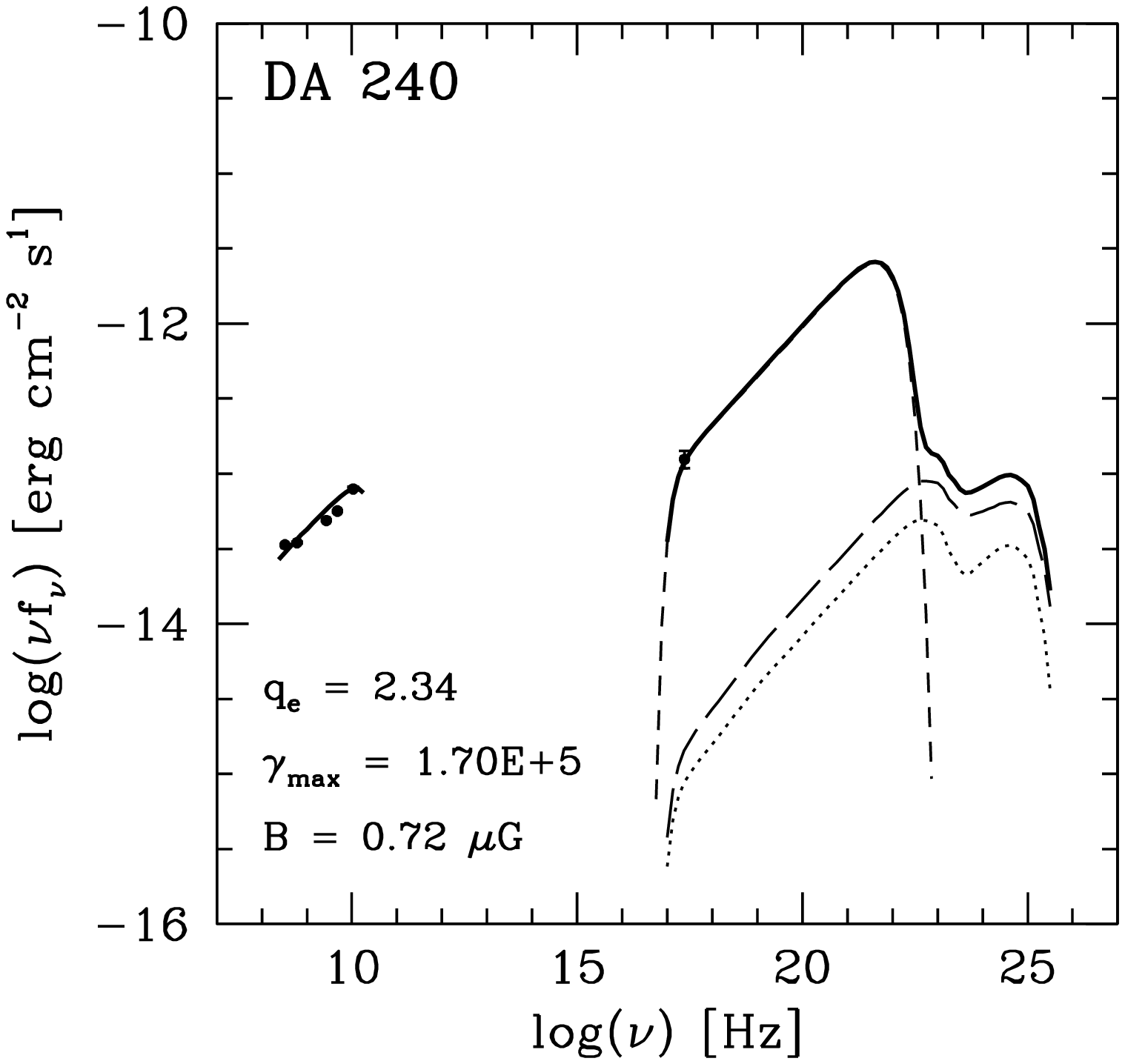}
\includegraphics{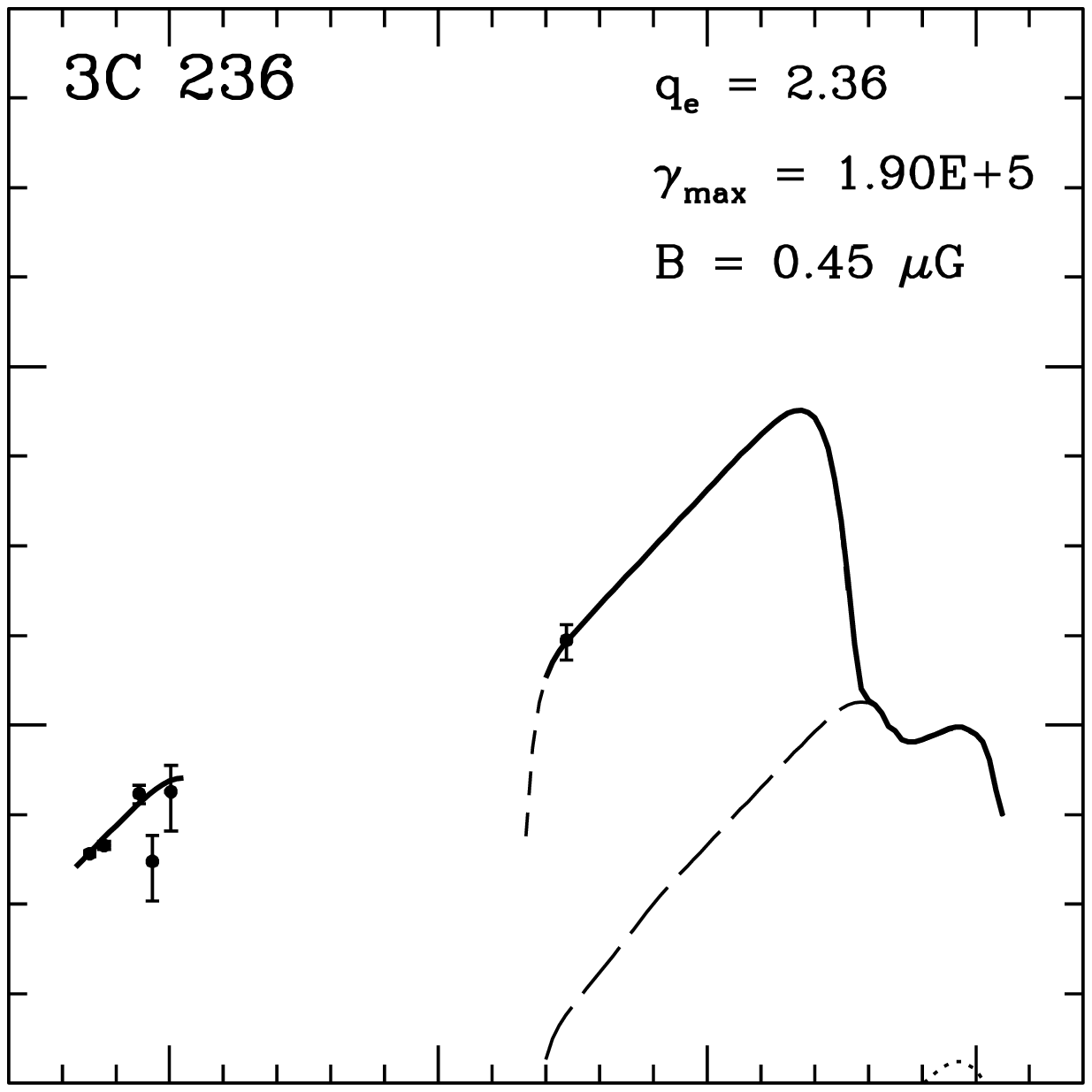}
\includegraphics{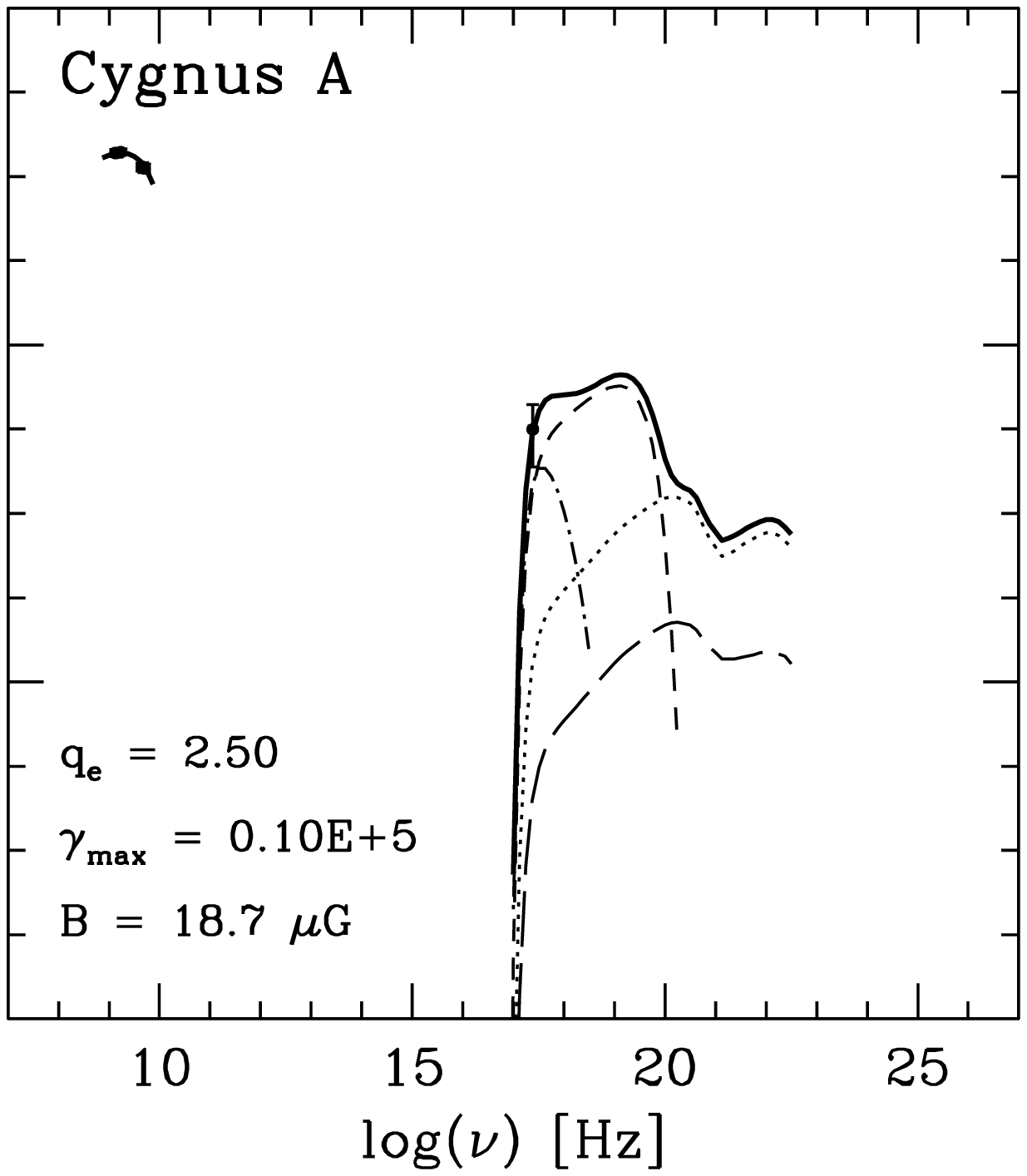}
\includegraphics{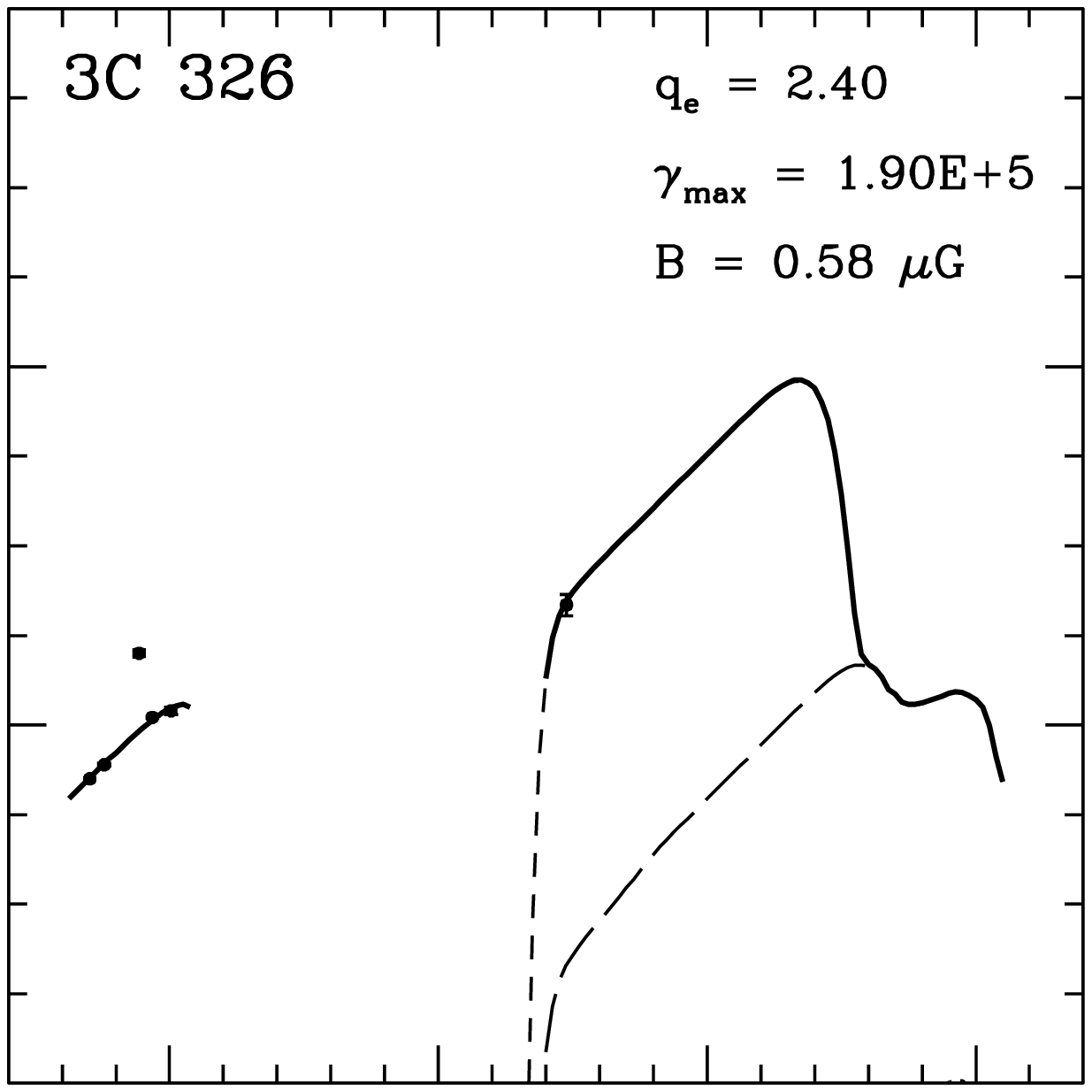}
\includegraphics{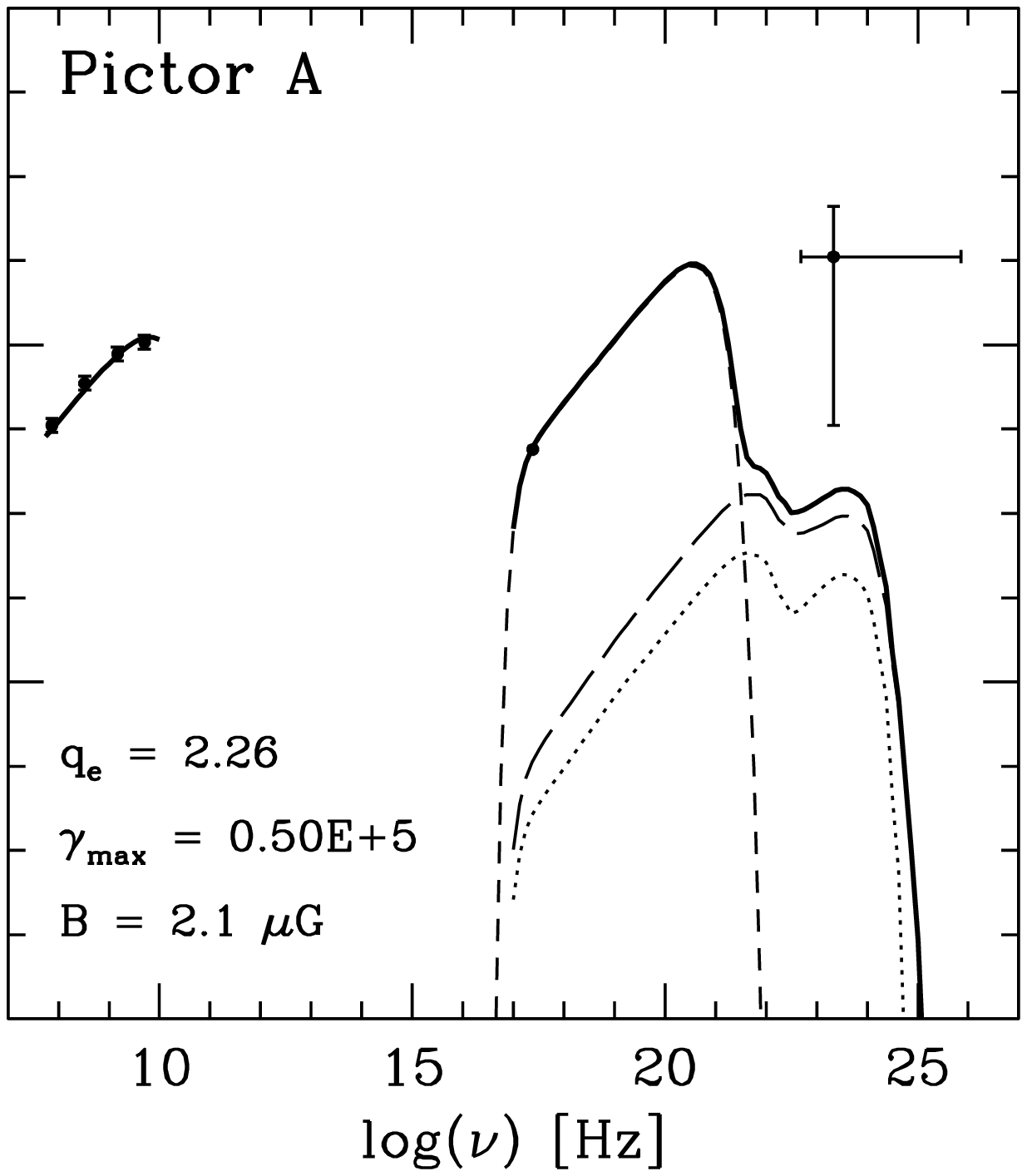}
\caption{
Data are denoted by dots (with error bars). Emission component curves are: 
synchrotron, solid; 
Compton/CMB, short-dashed;
SSC, dot-dashed (Cyg\,A); 
Compton/EBL, long-dashed; 
Compton/GFL, dotted; 
total Compton: thick solid. 
Indicated in each panel are the values of the EED parameters. 
For Cygnus\,A we show total lobe SED; the separate lobe SEDs are displayed in Fig.\,(\ref{fig:CygA_lobes_SEDs}). 
In Pictor\,A's SED we also show the lowest detected $\gamma$-ray flux (Brown \& Adams 2012). Upper limits to IR 
luminosities (see Section 2.2) are treated as nominal IR luminosities.
}
\label{fig:six_source_SED}
\end{figure*}

%
\begin{figure}
\vspace{12.0cm}
\includegraphics{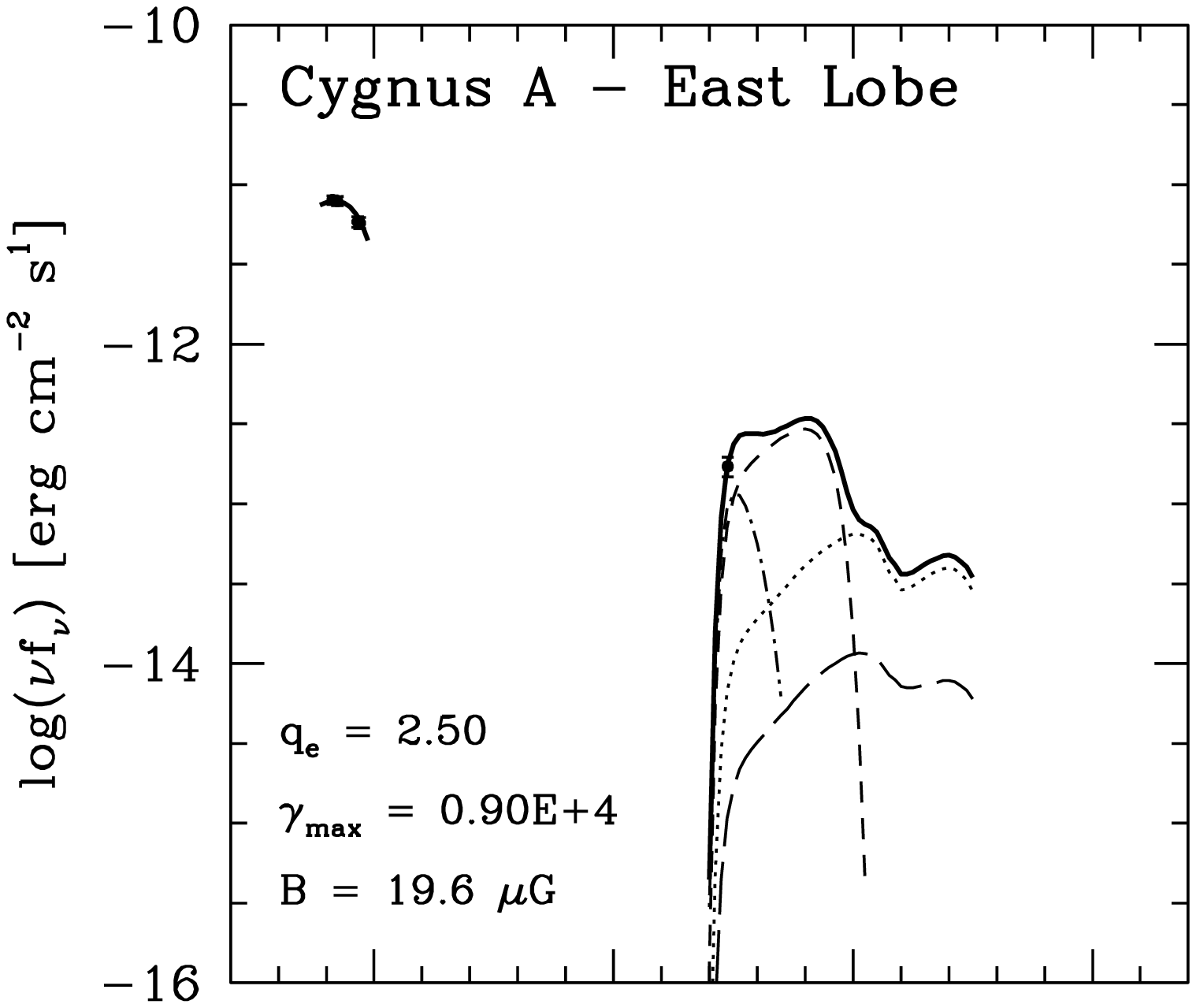}
\includegraphics{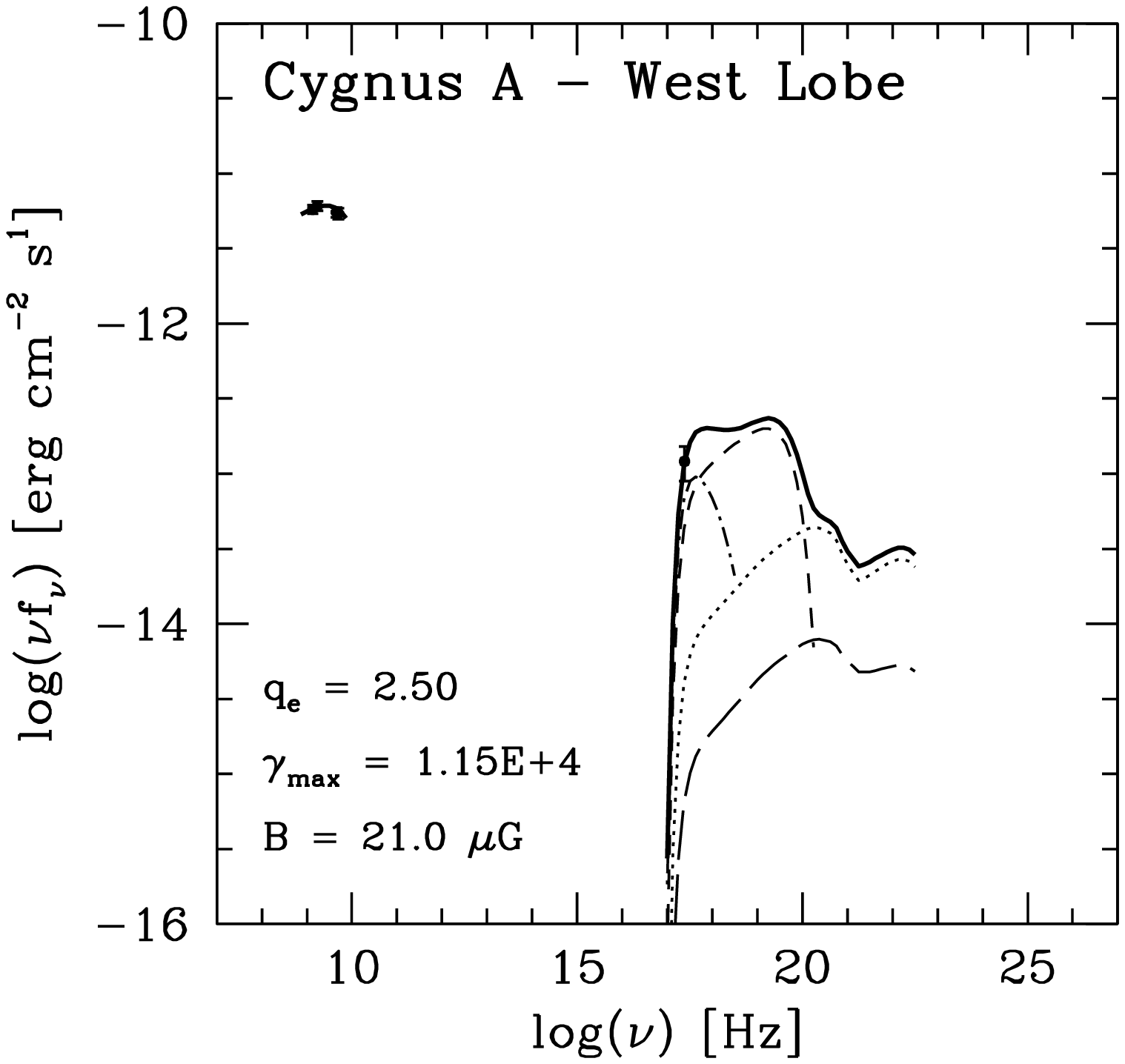}
\caption{ 
The SEDs of the E lobe (top) and the W lobe (bottom) of Cyg\,A. Symbols are as in Fig.\.(\ref{fig:six_source_SED}). 
}
\label{fig:CygA_lobes_SEDs}
\end{figure}
%

As is apparent from Fig.\,1, the radiative (synchrotron and Compton) yields of energetic electrons in the 
lobe SEDs analyzed here account for the currently available radio and X-ray data. This basic result of our 
spectral analyses strengthens a similar conclusion reached in earlier analyses (see Section 2.1).

\section{Discussion} 

The main conclusions of our analysis of the lobe SEDs of 3C\,98, 3C\,236, 3C\,326, DA\,240, Cygnus\,A, and 
Pictor\,A are quite similar to those presented in earlier papers where the original data for these sources 
were reported (see Section 2.1). However our treatment, first presented in Paper I, differs appreciably 
from those in previous analyses. We provide model SEDs for all sources using one simple EED (i.e., a single 
truncated PL) -- whereas previous studies did not explicitly model the SED (3C\,98, Isobe et al. 2005; 
3C\,236, Isobe \& Koyama 2015; 3C\,326, Isobe et al. 2009; DA\,240, Isobe et al. 2011), or did so using 
a broken PL (Cygnus\,A, Yaji et al. 2010 and deVries et al. 2018; Pictor\,A, Hardcastle \& Croston 2005). 
Our essentially uniform treatment allows us to compare the radiative properties of different lobes more 
directly and in a less constrained way. 

The electron energy density is determined by integration over their spectral distribution with PL index 
$q_e-1$ at energies below the characteristic value at which Coulomb losses are roughly equal to radiative 
losses, $\gamma < \gamma_{min}$, and with index $q_e$ for $\gamma > \gamma_{min}$ (e.g., Rephaeli \& Persic 
2015). In Table 7 we list estimated values of the electron and magnetic energy densities for the lobes 
under study. The electron to magnetic energy density ratios are in the range ${\mathcal O}$(10-100), suggesting a 
particle-dominated NT energy budget in the lobes. As mentioned, we cannot directly constrain proton contents 
in these lobes (unlike the case of Centaurus\,A, see Paper II), however the energy density of NT protons is 
probably much higher than that of NT electrons (see Paper II and Persic \& Rephaeli 2014). If so, this 
further substantiates the validity of the result that most of the lobe NT energy density is in energetic 
electrons and protons.

Our analysis indicates that a GFL contribution to the predicted Compton yields in the {\it Fermi}/LAT band 
\footnote{
We use projected galaxy-to-lobe distances (inclinations are unknown), so inferred GFL densities are strict upper limits.
}
is 
negligible in 3C\,236 and 3C\,326, dominant in Cyg\,A, and comparable to the EBL contribution in 3C\,98, 
DA\,240, and Pic\,A. The case of Cyg\,A is similar to that of Fornax\,A; in the lobes of both systems the 
dominant radiation field is the GFL, not the EBL. Note that  $\gamma$-ray emission from the Fornax\,A lobes 
was measured by {\it Fermi}/LAT (see Paper I). 

The lobes of Cyg\,A appear to have somewhat different SEDs, as was already noted (Yaji et al. 2010). In 
our simple truncated-PL characterization the difference is described by different EED high-energy cutoffs 
and magnetic fields. The radio spectrum shows a high-end spectral turnover at lower frequency in the E lobe 
than in the W lobe (Fig.\,\ref{fig:CygA_lobes_SEDs}). The shortest electron lifetime in the lobes, $\tau = 
10^{11}\, (\frac{4}{3} \frac{\sigma_T}{m_ec} \gamma_{max}$ $\frac{u_{\rm mag}+ u_{\rm syn} + u_{\rm CMB}}
{u_{\rm CMB}})^{-1}$ s (for the assumed isotropic electron distribution), is 0.26 Myr in the E lobe ($u_{\rm 
syn}^{\rm E} = 0.75$ eV cm$^{-3}$) and 0.18 Myr in the W lobe ($u_{\rm syn}^{\rm W} = 0.37$ eV cm$^{-3}$). 
If the electron spectrum has attained a steady-state, possibly by {\it in situ} re-acceleration, then the 
different lifetimes would imply that the process is more efficient in the W lobe.

\begin{figure}
\vspace{6.0cm}
\includegraphics{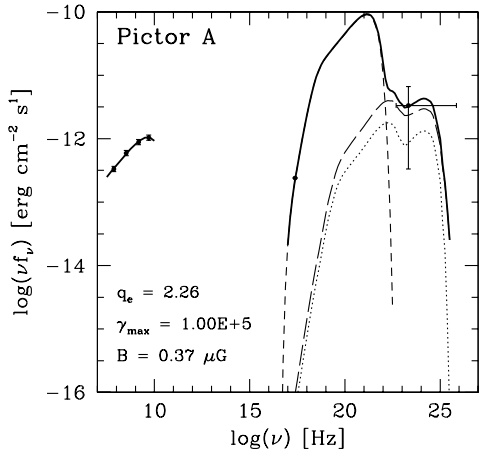}
\caption{
The SED of Pic\,A assuming the lowest $\gamma$-ray flux measured by {\it Fermi}/LAT (Brown \& 
Adams 2012) is diffuse lobe emission. Symbols are as in Fig.\.(\ref{fig:six_source_SED}). 
}
\label{fig:PicA_gamma_SED}
\end{figure}
%

The $\gamma$-ray emission of Pic\,A oberved by {\it Fermi}/LAT is unlikely to be related to the radio and 
X-ray emissions of a prominent compact hotspot in the Western lobe because a synchrotron self-Compton SED 
model linking the emissions in all three bands proves impossible (Brown \& Adams 2012). Indeed, the possibility 
that this $\gamma$-ray emission has a diffuse origin in the lobes can essentially be ruled out based on the 
following considerations: Lobe origin would require an EED normalization, $N_{e0}$, higher by an order of 
magnitude than in the no-$\gamma$-ray case. Such a high value of $N_{e0}$ would require an unusually high 
low-energy cutoff, $\gamma_{min} = 2500$, to account for the 1\,keV flux, and obviously a low magnetic field, 
$B = 0.37\,\mu$G, to fit the radio spectrum (see Fig.\,\ref{fig:PicA_gamma_SED}). The combination of high 
$N_{e0}$ and low $B$ would imply $u_e/u_B  \sim 1.8 \cdot 10^4$, an unusually high value in radio lobes (see 
Table 7, and Table 1 of Paper II). 

The predicted Compton/(EBL+GFL) $\gamma$-ray emission in the {\it Fermi}/LAT band is well below the LAT 
sensitivity, $\sim 10^{-12}$ erg s$^{-1}$ at 1\,GeV (5$\sigma$ detection for 10\,yr sky-survey exposure 
at high Galactic latitude; Principe et al. 2018) in all cases considered here. The average $\gamma$-ray 
to radio flux ratio, $\phi$, at their respective peaks (in $\nu f_\nu$ units) is $\phi \sim 1$ for 3C\,236, 
3C\,326, and DA\,240, $\phi \sim 0.1$ for 3C\,98 and Pic\,A, and $\phi \sim 0.01$ for Cyg\,A; and it is 
$\phi \sim 1$ for Fornax\,A, Cen\,A, Cen\,B, and $\phi \sim 100$ for NGC\,6251 (Paper II). Low $\phi$-ratios 
are a consequence of low $\gamma_{max}$: for example Cyg\,A, with $\gamma_{max} = 10^4$ (lowest value in the 
set), has the lowest $\phi$, whereas NGC\,6251, with $\gamma_{max}= 1.1 \cdot 10^6$ (highest value in the 
sample), has the highest-$\phi$; and other sources (e.g., Fornax\,A, Cen\,A) with intermediate values 
($\gamma_{max} \sim 10^5$) have $\phi \sim 1$. If, hypothetically, the radio data for Cyg\,A required $\gamma_
{max} \sim 10^5$, the predicted $\gamma$-ray flux would be $\magcir 10^{-12}$ erg s$^{-1}$ at 1\,GeV, detectable 
with {\it Fermi}/LAT. Thus, it appears that, apart from the obvious geometrical dimming, the predicted very low  
$\gamma$-ray fluxes reflect intrinsic properties of the electron populations in the halos. 

\begin{figure}
\vspace{6.0cm}
\includegraphics{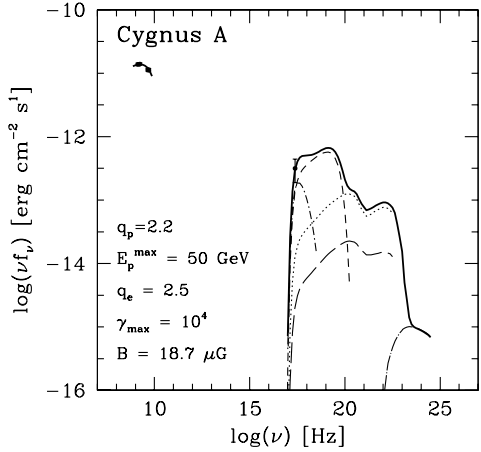}
\caption{
The lepto-hadronic model of the total lobe SED of Cyg\,A. 
Data are denoted by dots (with error bars). Emission component curves are: 
synchrotron, solid; 
Compton/CMB, short-dashed;
SSC, dot--short-dashed; 
Compton/EBL, long-dashed; 
Compton/GFL, dotted; 
pionic, dotted--long-dashed; 
total Compton and pionic: thick solid. 
Indicated are the values of the proton and electron spectral parameters. 
}
\label{fig:CygA_leptohadr}
\end{figure}

Cyg\,A is of particular interest with regard to the feasibility of $\gamma$-ray detection. Due to the low 
value of $\gamma_{max}$ the Compton/(EBL+GFL) hump cuts off at log$(\nu) \sim 22.5$, i.e. $\sim$100 MeV, 
just short of the {\it Fermi}/LAT range. 
If so, we would expect any detectable emission above this energy to be of pionic origin. 
This situation, along with the relatively high value of the thermal gas density in the lobes (Yaji et al. 2010)
\footnote{
The lack of internal depolarization in the lobes implies $n_H < 2 \cdot 10^{-4}$ cm$^{-3}$, assuming no field 
reversals in the lobes (Dreher et al. 1987). This results is clearly inconsistent, by 2 orders of magnitude, with the 
result of Yaji et al. (2010) who fit the X-ray spectra from the Cyg\,A lobes with a thermal + PL 
model. As suggested by Yaji et al., the discrepancy 
could be explained by a tangled field morphology.}, 
make the lobes of Cyg\,A interesting sources of pion-decay $\gamma$\,rays. The issue, of course, is 
the level of such emission.
Consider a proton energy distribution, $N(E_p) = N_{p0} (E_p/{\rm GeV})^{-q_p}$ with $q_p = 2.2$ and $E_p^{max} = 
50$ GeV. The former value is suggested by considerations of a moderately strong non-relativistic acceleration 
mechanism ($q_p = (R+2)/(R-1)$, with compression ratio $R \sim 3.5$), and of relatively low proton energy losses 
in the lobe environment (cf. the inelastic $pp$-interaction cross section in the relevant energy range, $\sigma_{pp} 
\sim 30$ mbarn; e.g. Kelner et al. 2006). The value of $E_p^{max}$ is not crucial; lower (higher) values in the range 
$\sim$5-100 GeV just make the pion hump somewhat narrower (broader). 
The energy density of the relativistic electrons, 1440 eV cm$^{-3}$, corresponds to a relativistic electron pressure 
of $p_e = u_e/3 = 7.7 \cdot 10^{-10}$ dyn cm$^{-2}$. To this we add the thermal pressure in the lobes, $p_{th} = n_H 
k_B T = 0.7 \cdot 10^{-10}$ dyn cm$^{-2}$ (from $n_H \sim 1.4 \cdot 10^{-2}$ cm$^{-2}$ and $k_B T=5.11$ keV; Yaji et 
al. 2010). As magnetic pressure is negligible, the (spectrally derived) total pressure is $p = p_e + p_{th} = 8.4 \cdot 
10^{-10}$ dyn cm$^{-2}$. The latter value may be compared with the pressure, $p = 8.6 \cdot 10^{-10}$ dyn cm$^{-2}$ 
(Snios et al. 2018), in the surrounding  "X-ray cocoon" (a.k.a. "X-ray cavity": Wilson et al 2006) that envelopes Cyg\,A. 
Assuming the lobes to be in pressure equilibrium with the cocoon (e.g. Mathews \& Guo 2010), we derive a NT proton 
pressure $p_p = 0.2 \cdot 10^{-10}$ dyn cm$^{-2}$: the corresponding energy density is $u_p = 3 p_p = 30$ eV cm$^{-3}$. 
Hence, $N_{p0} = 1.1 \cdot 10^{-8}$ cm$^{-3}$. The derived low proton content suggests that the relativistic fluid in the 
Cyg\,A lobes consists mainly of electron pairs, rather than a relativistic electron-proton plasma (e.g., Mathews 2014). 
This would indicate that energetic particles were transferred to the lobes by a "light" jet whose matter component 
consists largely of pair plasma (e.g., English et al. 2016; Snios et al. 2018). The resulting lepto-hadronic model is 
plotted in Fig.\,\ref{fig:CygA_leptohadr}, which clearly demonstrates that, even if weak, $\pi^0$-decay emission is 
likely to dominate emission in the {\it Fermi}/LAT band. 
%
%

\begin{table}
\caption[] {Energy densities (eV\,cm$^{-3}$) in the lobes.}

\begin{tabular}{ c  c  c  c  c  c  c  }
\hline
\hline

$u_{e,B}$         & 3C\,98 & 3C\,236 & 3C\,326 & DA\,240 & Cyg\,A & Pic\,A \\
\hline
$u_e$             & 19.5   & 0.04   & 0.38  & 0.11   & 1440   & 8.2   \\ 
$u_B$             & 0.17  & 0.005   & 0.008   & 0.013   & 8.70   & 0.11  \\ 
$\frac{u_e}{u_B}$ & 115   &  8 &  47.5   &  8.5   & 165.5  & 74.3   \\ 
\hline

\end{tabular}
\end{table}

\section{Conclusion}

Spectral distributions of NT electrons in lobes of radio galaxies can be directly determined from 
radio and X-ray measurements. Only weak upper limits can be set on energetic protons when there are 
no $\gamma$-ray measurements of the lobes.  

Modeling radio and X-ray measurements from several such lobes (3C\,98, Pictor\,A, DA\,240, Cygnus\,A, 
3C\,326, and 3C\,236 -- located at $D_L>125$ Mpc) as having synchrotron and Compton origin, we fully 
determine the spectral properties of the emitting electrons.

The main conclusion of our present analysis is that when NT X-ray emission measured for these sources is 
interpreted as Compton/CMB (including SSC, in the case of Cyg\,A) radiation, the ensuing SED (leptonic) 
models are similar to those for sources whose observational SEDs extend to the {\it Fermi}/LAT $\gamma
$-ray band (see Papers I and II). 

We do confirm earlier suggestions on the Compton/CMB nature of the diffuse 
X-ray emission from the original data papers. However, our treatment differs from those in previous analyses: 
we model all SEDs using a truncated single-PL EED, whereas the earlier studies either did not explicitly 
model the SED or did so using a broken PL. Our uniform treatment allows us to compare properties of 
different lobe SEDs in a more unbiased and direct way. 

We predict the Compton/(EBL+GFL) emission in the lobes using a recent EBL model (Franceschini \& Rodighiero 
2017; Acciari et al. 2019) and by accounting for the host galaxy contribution (GFL) to the superposed radiation 
fields in the lobes. Very low Compton fluxes in the {\it Fermi}/LAT band are predicted from sources whose radio 
spectra imply EEDs with low $\gamma_{max}$. 

For Cyg\,A we predict the Compton emission to be negligible at energies $\magcir$100 MeV, so any detectable 
emission in this spectral band would be of pionic origin and may allow a direct determination of the NT 
proton content.

\section*{Acknowledgements}
We used the NASA/IPAC Extragalactic Database (NED), which is operated by the Jet Propulsion Laboratory, 
Caltech, under contract with NASA. We acknowledge useful comments by an anonymous referee.

\end{document}